\documentclass[preprint,aps,prc,showpacs,nofootinbib]{revtex4-1}
\usepackage{amsmath}
\usepackage{amssymb}
\usepackage{graphics}
\usepackage{epsfig}

\usepackage{subfigure}
\begin{document}
\title{Manifestation of heavy  3/2-spin lepton
in large-angle $e^+e^-\rightarrow \gamma\gamma $ reaction at high energies}

\setlength{\fboxrule}{0.1em}
\author{ \fbox{G. I. Gakh}}

\affiliation{\it National Science Centre, Kharkov Institute of
Physics and Technology, Akademicheskaya 1, and V. N. Karazin
Kharkov National University, Dept. of
Physics and Technology, 31 Kurchatov, 61108 Kharkov, Ukraine}
\author{M.I. Konchatnij}
 \email{konchatnij@kipt.kharkov.ua}

 \affiliation{\it National Science Centre, Kharkov Institute of
Physics and Technology, Akademicheskaya 1, and V. N. Karazin
Kharkov National University, Dept. of
Physics and Technology, 31 Kurchatov, 61108 Kharkov, Ukraine}
\author{N.P. Merenkov}
\email{merenkov@@kipt.kharkov.ua}
\affiliation{\it National Science Centre, Kharkov Institute of
Physics and Technology, Akademicheskaya 1, and V. N. Karazin
Kharkov
National University, Dept. of
Physics and Technology, 31 Kurchatov, 61108 Kharkov, Ukraine}

\author{A. G. Gakh}
\email{agakh@karazin.ua}

\affiliation{V. N. Karazin
Kharkov National University, Dept. of
Physics and Technology, 31 Kurchatov, 61108 Kharkov, Ukraine}



\begin{abstract}
Manifestation of the heavy 3/2-spin lepton $(h^\pm)$, as possible virtual intermediate state in Feynmann diagrams, have been searched in the  $e^+\,e^-\to \gamma\,\gamma$
reaction at high energies and large photon angles. The spin-vector field $\Psi_\alpha$ of the 3/2-lepton is described by the Rarita-Schwinger one and
phenomenological Lagrangian of $h^\pm\, e^\pm \, \gamma$-interaction is chosen similarly to interaction of $\Delta$ isobar with nucleon and $\gamma$ quant.
It is described by two constants with dimension $[M^{-1}]$ and $[M^{-2}].$
The differential cross section and polarization asymmetries have been calculated for the case when both beams are polarized longitudinally along their directions, as well transversally, in the reaction plane, and normally, perpendicularly to it. Numerical estimations are performed in wide diapason of
the collision energy and parameters entering phenomenological Lagrangian.
\end{abstract}

\vspace{0.2cm}
\vspace{0.2cm}

\maketitle

\section{Introduction}

Recent analysis of experimental data collected from 1989 up to 2003, to search the non-pointness of the electron in $e^+\,e^-$ annihilation reactions (see \cite{Chen:2022jgm} and references therein) showed a significant signal at a confidence level 5 standard deviations. To explain this finding authors suggest the existence of an exited electron with mass about 310\,GeV and some contact interaction with a cutoff scale about 1250\,GeV. Although the more recent experimental data obtained in electron-proton \cite{H1:2008jzo} and  proton-proton \cite{D0:2008gqz, ATLAS:2013brl, CMS:2015lgt} collisions do not see an evidence for the exited lepton, this result is awaiting confirmation at future circular and linear electron-positron colliders  such as FCC \cite{FCC:2018byv}, CEPC \cite{CEPCStudyGroup:2018ghi}, ILC \cite{Behnke:2013xla} and ClIC \cite{Aicheler:2012bya}.

In present paper we consider one more mechanism that can influence such deviation (it may be more exotic one): effects due to existence of a heavy lepton $h^\pm$ with spin 3/2 which, like a   heavy exited electron, appears as a virtual intermediate state in Feynman diarams. We assume that 3/2-spin heavy lepton is described by the Rarita-Schwinger spin-vector field \cite{PhysRev.60.61}
and it interacts with  at least the QED sector of the Standard Model and can decay into  electron and $\gamma$ quant just similarly to decay of the $\Delta$ isobar into nucleon and photon (see, for example, \cite{Jones:1972ky, Pascalutsa:2006up}). Therefore it is quite natural to look for the manifestation of this lepton in the large-angle high energy annihilation reaction
\begin{equation}\label{eq:reaction}
e^-(p_1)+e^+(p_2)\to \gamma(k_1)+\gamma(k_2).
\end{equation}

The possibility of the existence 3/2-spin heavy lepton within composite models \cite{Harari:1982xy} and searching for a signal associated with it have been discussed early in a number works \cite{LeiteLopes:1980mh, Fleury:1982pp, Choudhury:1984bu, Almeida:1995yp, Cakir:2007wn}. Here we mainly focus on the effects due to arbitrary beam polarizations and consider the influence of the virtual  intermediate heavy lepton on various polarization asymmetries in the angular distribution of photons.

\section{Formalism}

To estimate effects due to 3/2-spin heavy lepton in the reaction (\ref{eq:reaction}) we restrict ourselves with standard QED diagrams (Fig.~1\,a)) and diagrams with intermediate
heavy lepton $h^\pm$ (Fig.~1\,b)) instead of electron (HL diagrams). We consider the case when both, electron and positron, can be arbitrary polarized, namely: longitudinally (along their 3-momentum direction) with 4-vector $S_{1\mu}^l$, transversally (in the reaction plane) with $S_{1\mu}^t$ and normal (perpendicularly to the reaction plane) with $S_{1\mu}^n$. Our aim is to search manifestation  of the HL diagrams on the level differential cross section and different polarization asymmetries at high beam energies achievable at future linear colliders ILC and CLIC.

\subsection{Normalization and polarization 4-vectors}

We choose a normalization such that, being averaged over the electron and positron polarizations, the differential angular cross section reads
\begin{equation}\label{eq:norm}
\frac{d\,\sigma}{d\,\Omega} = \frac{\alpha^2}{16\,s}|M^2|, \ s=(p_1+p_2)^2, \ t=(p_1-k_1)^2, \ u=(p_1-k_2)^2,
\end{equation}
where $\alpha =e^2/(4\pi) =1/137$ fine structure constant and the matrix element squared is defined by a sum of the QED, QED-HL interference and HL contributions.
The same expression is valid when electron and positron are polarized. When calculating traces in $|M^2|$ at chosen normalization we have to use the electron and positron projections as
(we omit the spin indecies)
\begin{equation}\label{eq:p1p2}
u(p_1)\,\bar{u}(p_1) = (\hat{p}_1+m)(1+\gamma_5\,\hat{S}_1), \ \ v(p_2)\,\bar{v}(p_2) = (\hat{p}_2-m)(1+\gamma_5\,\hat{S}_2),
\end{equation}
where $u(p_1)$  and $v(p_2)$ are the electron and positron bispinors, respectively.

At high energies, neglecting the electron mass, we can express 4-vectors $S_{1,2\mu}^{l,t,n}$  in terms of particle 4-momenta, namely
\begin{equation}\label{eq:S1S2}
S_{1\mu}^l = \lambda_1\frac{p_{1\mu}}{m}, \ S_{2\mu}^l =  \lambda_2\frac{p_{2\mu}}{m}, \ S_{1\mu}^t =
\eta_1\sqrt{\frac{s}{u\,t}}\Big(\frac{u}{s}p_{1\mu} + \frac{t}{s}p_{2\mu} + k_{1\mu}\Big), \ S_{2\mu}^t = \frac{\eta_2}{\eta_1}S_{1\mu}^t,
\end{equation}
$$S_{1\mu}^n=\xi_1\frac{\varepsilon_{\mu\nu\lambda\rho}\, p_1^\nu\, p_2^\lambda \,k_1^\rho}{K}, \ K=\frac{1}{2}\sqrt{stu}, \ S_{2\mu}^n= \frac{\xi_2}{\xi_1}\,S_{1\mu},$$
where $\lambda_{1,2},\,\eta_{1,2}$ and $\xi_{1,2}$ are corresponding polarization degrees. In the case of longitudinal polarization one must  first perform multiplication in (\ref{eq:p1p2}) and then go to the limit $m\to 0.$ Take attention that part of the differential cross section which does not depend on polarizations is the same for unpolarized and polarized beams.

\begin{figure}
\centering
\includegraphics[width=0.45\textwidth]{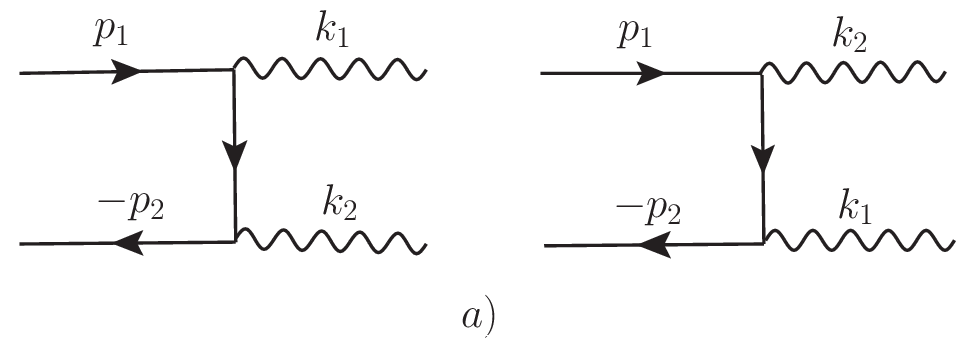}
\includegraphics[width=0.45\textwidth]{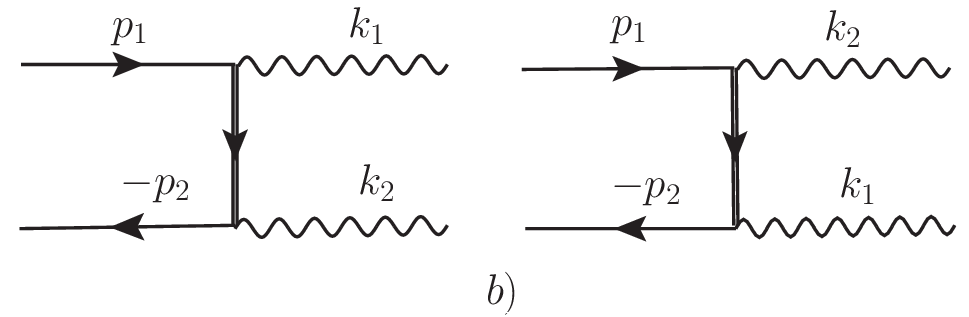}
 \parbox[t]{0.9\textwidth}{\caption{Feynman diagrams for the process $e^+\,e^-\rightarrow \gamma\,\gamma$. a)\,QED diagrams,
  b)\,diagrams with the heavy 3/2-spin lepton $h$ in $t-$ and $u-$ channels.}\label{fig.1}}
\end{figure}

Let $\big(d\sigma/d\Omega\big)_{un}$ is the cross section in unpolarized case and $\big(d\sigma/d\Omega\big)_{i\,j},\, i,j =l,\,t,\,n$ is the cross section in the case
of positron polarization $S_{2}^i$ and electron one $S_{1}^j$. Then we define the
polarization asymmetries $A_{ll},\,\,A_{tt},\,\,A_{nn},\,\,A_{lt}$ and $A_{tl}$  as follows

\begin{equation}\label{eq:Aij}
\frac{d\sigma_{ll}}{d\Omega} = \frac{d\sigma_{un}}{d\,\Omega}\big(1 \pm \lambda_1\,\lambda_2 A_{ll}\big),  \
\frac{d\sigma_{tt}}{d\Omega} = \frac{d\sigma_{un}}{d\,\Omega}\big(1 \pm \eta_1\,\eta_2 A_{tt}\big), \
\frac{d\sigma_{nn}}{d\Omega} = \frac{d\sigma_{un}}{d\,\Omega}\big(1 \pm \xi_1\,\xi_2 A_{nn}\big),
\end{equation}
$$\frac{d\sigma_{lt}}{d\Omega} = \frac{d\sigma_{un}}{d\,\Omega}\big(1 \pm \eta_1\,\lambda_2 A_{lt}\big),  \
\frac{d\sigma_{tl}}{d\Omega} = \frac{d\sigma_{un}}{d\,\Omega}\big(1 \pm \lambda_1\,\eta_2 A_{tl}\big). $$

Signs in front $A_{ij}$ can be changed by changing the sign of one of the 4-vector polarization to the opposite; sign "+" corresponds to choice given by definitions (\ref{eq:S1S2}).

Note that $A_{nt} = A_{nl} =A_{tn} = A_{ln} = 0$ for obvious reason. In the limit $m\to 0$ the Born approximation gives $A_{lt}^B = A_{tl}^B =0$ and
\begin{equation}\label{eq:Born}
\left(\frac{d\sigma}{d\Omega}\right)_{un}^B = \frac{\alpha^2}{s}\Big(\frac{1+\cos^2{\theta}}{1-\cos^2{\theta}}\Big), \ \ A_{ll}^B =-1,
\ A_{tt}^B = A_{nn}^B = \frac{1-\cos^2{\theta}}{1+\cos^2{\theta}}.
\end{equation}

\subsection{About 3/2-spin heavy lepton}

We suggest that even-parity 3/2-spin lepton $h^\pm$ is described by the Rarita-Schwinger spin-vector field $\Psi^\alpha_p (x) = u^\alpha (p)\exp(-ip\cdot x)$
and Lagrangian its interaction with electron and photon is very similar to the Lagrangian describing electromagnetic decay of
$\Delta $ isobar into nucleon and $\gamma$ quant: $\Delta \to N + \gamma.$ There are a lot possibilities to choose this Lagrangian, see for example, \cite{Pascalutsa:1998pw, Pascalutsa:1999zz, Kristiano:2017qjq}
but we will use well known one from \cite{Jones:1972ky}
\begin{equation}\label{eq:Lagr}
L_{\Psi^\alpha\,e\, \gamma} = e\,\bar{\psi}\,\gamma_5\Big(\frac{ig_1}{M}\gamma^\mu-\frac{g_2}{M^2}\partial^\mu\,\Big)\Psi^\alpha\,F_{\alpha\mu} +h.\,c.,
\ F_{\alpha\mu}=\partial_\mu\,A_\alpha-\partial_\alpha\,A_\mu,
\end{equation}
where dimensionless constants $g_1$ and $g_2$ are real (C\,P parity is conserved), parameter $M$ has the dimension of mass, and $F_{\alpha\mu}$ is tensor of an electromagnetic field.
Below we take mass of the $h^\pm$ as the parameter $M$  and assume $M^2 \approx s.$

Lagrangian (\ref{eq:Lagr}) defines an amplitude $A_D$ of the decay $h^-(p_1) \to e^-(p_2)+\gamma(k)$ in terms of constants $g_1$ and $g_2$ as follows
\begin{equation}\label{eq:Aeg}
A_D = e\,\bar{u}(p_2)\,\gamma_5\,\Gamma^{\alpha\mu} \,u_\alpha(p_1)\,\varepsilon^*_\mu(k),
\end{equation}
where
$$\Gamma^{\alpha\mu} = \frac{g_1}{M}(k^\alpha\,\gamma^\mu - \hat{k}\,g^{\alpha\mu}) +\frac{g_2}{M^2}[k^\alpha\,p_1^\mu -(k\cdot p_1)g^{\alpha\mu}].$$
The square of this amplitude reads
\begin{equation}\label{eq:AD2}
|A_D|^2 = e^2\,Sp\big\{(\hat{p}_2-m)\,\Gamma^{\alpha\mu}\,(\hat{p}_1+M)\,P_{\alpha\beta}(p_1)\Gamma^{\beta}_\mu\big\},
\end{equation}
$$P_{\alpha\beta}(p_1) = g_{\alpha\beta}-\frac{1}{3}\gamma_\alpha\,\gamma_\beta -\frac{1}{3\,p_1^2}\big(\hat{p}_1\gamma_\alpha p_{1\beta}
 +p_{1\alpha}\gamma_{\beta}\hat{p}_1\big), \ \ \Gamma^{\beta}_\mu = \Gamma^{\beta\lambda}\,g_{\lambda\mu},$$
where we taken into account that $\sum_{pol}\varepsilon^*_\mu\varepsilon_\rho = -g_{\mu\rho}.$
Using conservation low one obtains the partial decay width neglecting the electron mass
\begin{equation}\label{eq:DW}
\Gamma(h^\pm\to e^\pm\,\gamma) = \frac{\alpha\,M}{24}\big(3g_1^2 +g_2^2+3g_1\,g_2\big).
\end{equation}
Note that in this limit the analytic form of the corresponding negative-parity HL decay width coincides with (\ref{eq:DW}).

\subsection{Unpolarized cross section}

In accordance with Feynmann rules, contribution HL diagrams to the reaction amplitude can be written as follows
(we extracted factor $\alpha^2$ in accordance with the definition (\ref{eq:norm}) of the differential cross section)
\begin{equation}\label{eq:AmHL}
M_h =\bar{v}(p_2)\gamma_5\big[\Gamma^{\beta\nu}_{22}\Delta_{\beta\alpha}(p)\Gamma\,^{\alpha \mu}_{11}
+\Gamma^{\beta\mu}_{21}\Delta_{\beta\alpha}(q)\Gamma^{\alpha\nu}_{12}\big]\gamma_5 u(p_1)\epsilon^*_\mu(k_1)\,\epsilon^*_\nu(k_2),
\end{equation}
where $\epsilon_\mu(k_1)$ and $\epsilon_\nu(k_2)$ are 4-vectors of photon polarizations and
$$p=p_1-k_1=k_2-p_2, \ q=p_1-k_2=k_1-p_2.$$
The sort notation for all $\Gamma$'s in (\ref{eq:AmHL}) is
\begin{equation}\label{eq:shortG}
\Gamma^{\beta\nu}_{ij}=\frac{g_1}{M}(k_j^\beta\gamma^\nu - \hat{k}_j\,g^{\beta\nu}) + \frac{g_2}{M^2}[k_j^\beta p_i^\nu-(p_i\cdot k_j)\,g^{\beta\nu}], \, (i,j =1,2)
\end{equation}
Propagator $\Delta_{\alpha\beta}$ of the particle with spin 3/2 reads (see Eqs.\,(13),\,(14) in Ref.\,\cite{PT1999})
\begin{equation}\label{eq:propD}
\Delta_{\alpha\beta}(p)=-\frac{\hat{p}+M}{p^2-M^2}P_{\alpha\beta}(p), \ P_{\alpha\beta}(p)\,p^\beta = p^\alpha\,P_{\alpha\beta}(p) =0, \ P_{\alpha\beta}(p)\,\gamma^\beta = \gamma^\alpha\,P_{\alpha\beta}(p) =0.
\end{equation}

To obtain $|M_h|^2$ one has to compute rather long trace
\begin{equation}\label{eq:MhSpur}
|M_h|^2 = Spur\big\{(\hat{p}_2-m)\gamma_5\big[\Gamma^{\beta\nu}_{22}\Delta_{\beta\alpha}(p)\Gamma\,^{\alpha \mu}_{11}
+\Gamma^{\beta\mu}_{21}\Delta_{\beta\alpha}(q)\Gamma^{\alpha\nu}_{12}\big]\gamma_5\times
\end{equation}
$$(\hat{p_1}+m)\gamma_5\big[\Gamma_{11\, \mu}^\delta\widetilde{\Delta}_{\delta\sigma}(p)\Gamma_{22\,\nu}^\sigma
+ \Gamma_{12\, \nu}^\delta\widetilde{\Delta}_{\delta\sigma}(q)\Gamma_{21\, \mu}^\sigma\big]\gamma_5\big\},
\ \widetilde{\Delta}_{\delta\sigma}(p) =- P_{\delta\sigma}(p)\frac{\hat{p}+M}{p^2-M^2}.$$

As concerns the interference of the pure QED and HL diagrams, it can be written as
\begin{equation}\label{eq:MheSpur}
2\,M_h\,M_e^{*} = 2\,Spur\big\{(\hat{p}_2-m)\gamma_5\big[\Gamma^{\beta\nu}_{22}\Delta_{\beta\alpha}(p)\Gamma\,^{\alpha \mu}_{11}
+\Gamma^{\beta\mu}_{21}\Delta_{\beta\alpha}(q)\Gamma^{\alpha\nu}_{12}\big]\gamma_5\times
\end{equation}
$$(\hat{p}_1+m)\big[\gamma_\mu\frac{\hat{p}+m}{p^2-m^2}\gamma_\nu + \gamma_\nu\frac{\hat{q}+m}{q^2-m^2}\gamma_\mu \big] \big\}.$$

In what follows we neglect the electron mass and introduce the dimensionless invariant variables
\begin{equation}\label{eq:DlIV}
s=\frac{2(p_1\cdot p_2)}{M^2}, \ t= -\frac{2(p_1\cdot k_1)}{M^2}, \ u=-\frac{2(p_1\cdot k_2)}{M^2}, \ s+t+u =0, \ \cos{\theta}= 1+\frac{2 t}{s}.
\end{equation}

These variables are very convenient especially  in the limit $m\to 0.$ In terms of these variables we obtain for HL diagrams contribution into matrix element squared
\begin{equation}\label{eq:Mh2}
\left|M_h\right|^2 =2 g_1^4\, A_{11}(s,t)+\frac{1}{3}g_1^2\,g_2^2\, A_{12}(s,t) + \frac{1}{36}g_2^4\, A_{22}(s,t),
\end{equation}
where
$$A_{11}(s,t)=-2[3+2s^2+t^2+s(5+t)]-(1+3s+3s^2)\Big[\frac{1}{(1-t)^2}+\frac{1}{(1+s+t)^2}\Big]+$$
$$\frac{8+22s+21s^2+10s^3+s^4}{2+s}\Big(\frac{1}{1-t}+\frac{1}{1+s+t}\Big),$$
$$A_{12}(s,t)=4s^3+3s^2(11+6t)+12(2+t)^2 +6s(11+2t+3t^2)+3(1+2s)^2\Big[\frac{1}{(1-t)^2}+\frac{1}{(1+s+t)^2}\Big]$$
$$-\frac{30+114s+126s^2+46s^3+s^4}{2+s}\Big(\frac{1}{1-t}+\frac{1}{1+s+t}\Big),$$
$$A_{22}(s,t)=s^3(23-5t)-s^2(74+44t+21t^2)-4s(21+9t+11t^2+8t^3)-4(14+9t^2+4t^4)$$
$$-(5+6s+6s^2)\Big[\frac{1}{(1-t)^2}+\frac{1}{(1+s+t)^2}\Big]$$
$$+\frac{66+152s+175s^2+52s^3-18s^4}{2+s}\Big(\frac{1}{1-t}+\frac{1}{1+s+t}\Big).$$

The QED-HL interference contributes as follows
\begin{equation}\label{eq:MhMe*}
2\,Re[M_h\,M_e^{*}]=8\,g_1^2\Big[\frac{(1+s)^2}{1-t} + \frac{(1+s)^2}{1+s+t} -2-3s\Big]+
\end{equation}
$$\frac{2\,g_2^2}{3}\Big[8+14s+s^2+8st+8t^2 - (4+9s+3s^2)\Big(\frac{1}{1-t}+\frac{1}{1+s+t}\Big)\Big].$$

To calculate the angular distribution in high energy limit one can use
$$A_{11}(s,{\theta})= \frac{4(8+22s+21s^2+10s^3+s^4)}{(2+s)^2-s^2\cos^2{\theta}} -6-10s-\frac{s^2(7+\cos^2{\theta})}{2}$$
$$-4(1+3s+3s^2)\Big[\frac{1}{(2+s-s\cos{\theta})^2} + \frac{1}{(2+s+s\cos{\theta})^2}\Big],$$
$$A_{12}(s,{\theta})=24 + 66s +30s^2-\frac{s^3}{2}+3s^2\cos^2{\theta}\Big(1+\frac{3s}{2}\Big)$$
$$+12(1+2s)^2\Big[\frac{1}{(2+s-s\cos{\theta})^2} + \frac{1}{(2+s+s\cos{\theta})^2}\Big] -\frac{4[30+114s+126s^2+46s^3+s^4]}{(2+s)^2-s^2\cos^2{\theta}},$$
$$A_{22}(s,{\theta})=-56-84s-65s^2+34s^3+\frac{s^4}{4}-\Big(9+11s-\frac{3s^2}{4}\Big)s^2\cos^2{\theta}-s^4\cos^4{\theta}+$$
$$\frac{4(66+152s+175s^2+52s^3-18s^4)}{(2+s)^2-s^2\cos^2{\theta}}-(5+6s+6s^2)\Big[\frac{4}{(2+s-s\cos{\theta})^2} + \frac{4}{(2+s+s\cos{\theta})^2}\Big].$$

Regards the QED-HL interference we have
\begin{equation}\label{eq:MhMetheta}
2\,Re[M_{3/2}^+M^{e*}]=8e^2\,g_1^2\Big[-2-3s +\frac{4(1+s)^2(2+s))}{(2+s)^2-s^2\cos^2{\theta}}\Big] +
\end{equation}
$$\frac{2e^2\,g_2^2}{3}\Big[8+14s+s^2-2s^2\sin^2{\theta}-\frac{4(4+9s+3s^2)(2+s)}{(2+s)^2-s^2\cos^2{\theta}}\Big].$$
Note that in unpolarized case neither $|M_h|^2$ nor $Re[M_h\,M_e^{*}]$ contains product of the odd powers $g_1$ and $g_2,$ so they do not depend on relative sign of these constants.
With accounting for all the contribution, we can write down the the differential over the photon solid angle cross section in unpolarized case as
\begin{equation}\label{eq:dsun}
\frac{d\sigma_{un}}{d\,\Omega}=\frac{\alpha^2}{16 M^2 }\Big[\frac{16(1+\cos^2\theta)}{s(1-\cos^2\theta)}
+ \frac{2\,Re[M_h\,M_e^{*}]}{s} + \frac{|M_h|^2}{s}\Big].
\end{equation}

In Fig.\,2 we plot the differential cross section (\ref{eq:dsun}) (in units $M^{-2})$ at different values of constant $g_1: 1,\,0.35,\,0.1$ and two angles $\theta=45^o,\,90^o$
as a function of the ratio $g=g_2/g_1$ and dimensionless invariant variable $s$.

\begin{figure}
\centering
\includegraphics[width=0.3\textwidth]{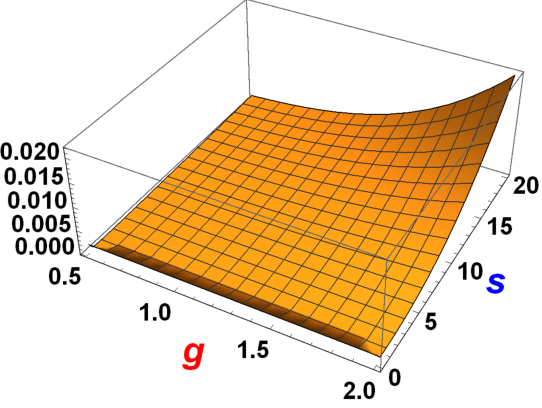}
\includegraphics[width=0.3\textwidth]{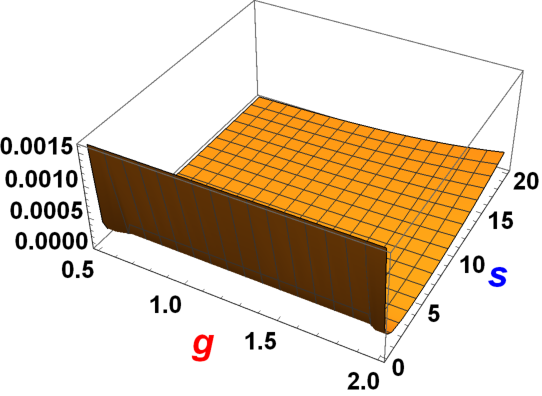}
\includegraphics[width=0.3\textwidth]{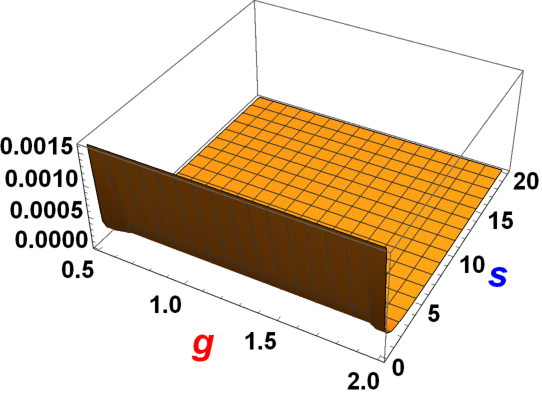}

\vspace{0.5cm}
\includegraphics[width=0.3\textwidth]{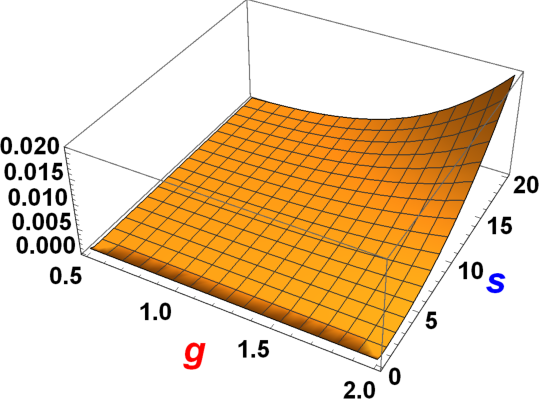}
\includegraphics[width=0.3\textwidth]{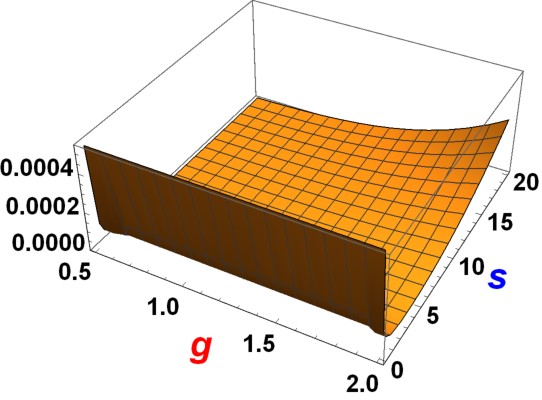}
\includegraphics[width=0.3\textwidth]{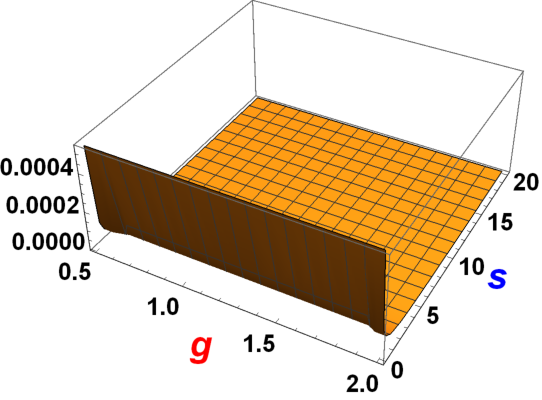}
 \parbox[t]{0.9\textwidth}{\caption{Unpolarized differential cross section defined by Eq.\,(\ref{eq:dsun}) (in units $M^{-2}$) at two values of the photon angle: $\theta=45^o,$ the upper row; $\theta=\,90^o$, the lower row; and three values of constants $g_1$: 1$-$left panel; 0.35$-$middle panel; 0.1$-$ right panel. $g$ is the ratio $g_2/g_1$.}\label{fig.2}}
\end{figure}

\section{Polarization asymmetries}

To estimate different polarization asymmetries one has to compute corresponding polarization dependent parts of the cross section, using the electron and positron projections as given by
Eq.(\ref{eq:p1p2}) and expressions for different polarization 4-vectors defined by relations (\ref{eq:S1S2}).

\subsection{Case of longitudinal polarization}

The direct calculation in the case of longitudinal polarization of the electron and positron along beam directions results
\begin{equation}\label{eq:Mhll}
\left|M_h^{+ll}\right|^2=2g_1^4L_{11}(s,t)+\frac{1}{3}L_{12}(s,t) +\frac{1}{36}L_{22}(s,t),
\end{equation}
$$L_{11}(s,t)=2[3+2s^2+t^2+s(5+t)]+(1+3s+3s^2 +2s^3)\Big[\frac{1}{(1-t)^2}+\frac{1}{(1+s+t)^2}\Big]$$
$$-\,\frac{8+22s+21s^2+6s^3+s^4}{2+s}\Big(\frac{1}{1-t}+\frac{1}{1+s+t}\Big),$$
$$L_{12}(s,t)=2s^3+3s^2(5+2t)-12(2+t^2)-6s(3+2t-t^2)-3(1+2s)\Big[\frac{1}{(1-t)^2}+\frac{1}{(1+s+t)^2}\Big]+$$
$$\frac{30+54s+6s^2-14s^3+s^4}{2+s}\Big(\frac{1}{1-t}+\frac{1}{1+s+t}\Big),$$
$$L_{22}(s,t)=s^3(47+5t) +s^2(178+44t+21t^2)+4s(43+9t+11t^2+8t^3)+4(14+9t^2+4t^4)+$$
$$(5+28s+54s^2+36s^3)\Big[\frac{1}{(1-t)^2}+\frac{1}{(1+s+t)^2}\Big]$$
$$-\, \frac{66+284s+463s^2+320s^3+78s^4}{2+s}\Big(\frac{1}{1-t}+\frac{1}{1+s+t}\Big).$$

For the corresponding contribution into QED HL interference we have
\begin{equation}\label{eq:MhMtll}
2\,Re[M_h\,M_e^{*}]^{ll}=-8\,g_1^2\frac{[s^3+2t^2+s^2(1+3t)+st(2+3t)]}{(1-t)(1+s+t)}+
\end{equation}
$$2\,g_2^2\frac{8t^4+s^3(2+t)+3s^2t(2+3t)+2st^2(3+8t)}{3(1-t)(1+s+t}.$$
As one can see, the odd powers of constant $g_1$ and $g_2$ are also absent as in the unpolarized case.

To study the $(s,\,\theta)$ distribution one can use for contribution into $\left|M_h^{+ll}\right|^2$
$$L_{11}(s,\theta)= 6+10s +4s^2 -\frac{s^2}{2}\,\sin^2{\theta}- \frac{4(8+22s+21s^2+6s^3+s^4)}{(2+s)^2-s^2\cos^2{\theta}}+$$
$$4(1+3s+3s^2+2s^3)\big[\frac{1}{(2+s-s\cos{\theta})^2} + \frac{1}{(2+s-s\cos{\theta})^2}\big],$$
$$L_{12}(s,\theta)=-24-18s+15s^2+23s^3 +3s^2\Big(1-\frac{s}{2}\Big)\sin^2{\theta} + \frac{4(30+54s+6s^2-14s^3+s^4)}{(2+s)^2-s^2\cos^2{\theta}}$$
$$-12(1+2s)\big[\frac{1}{(2+s-s\cos{\theta})^2} + \frac{1}{(2+s-s\cos{\theta})^2}\big],$$
$$L_{22}(s,\theta)=56+172s + 178s^2 +47s^3 +\Big(s^2\sin^2{\theta} -9-11s-\frac{5s^2}{4}\Big)s^2\sin^2{\theta}+$$
$$4(5+28s+54s^2+36s^3)\big[\frac{1}{(2+s-s\cos{\theta})^2} + \frac{1}{(2+s-s\cos{\theta})^2}\big]$$
$$-\,\frac{4(66+284s+463s^2+320s^3+78s^4)}{(2+s)^2-s^2\cos^2{\theta}}.$$

The corresponding result for the QED-HL interference reads
\begin{equation}\label{eq:MhMelltheta}
2\,Re[M_h\,M_e^{*}]^{ll}=-8\,g_1^2 s^2\frac{4(1+s)-(2+3s)\sin^2{\theta}}{(2+s)^2-s^2\cos^2{\theta}}+ \frac{2}{3}\,g_2^2\,s^3\frac{8+(2s\sin^2{\theta}-6-s)\sin^2{\theta}}{(2+s)^2-s^2\cos^2{\theta}}.
\end{equation}

Asymmetry $A_{ll}$, in the case of longitudinally polarized electron and positron along their beam
directions, is plotted In Fig.\,3 at the same values of the angle $\theta$ and constant $g_1$ as in Fig.\,2.

\begin{figure}
\centering
\includegraphics[width=0.3\textwidth]{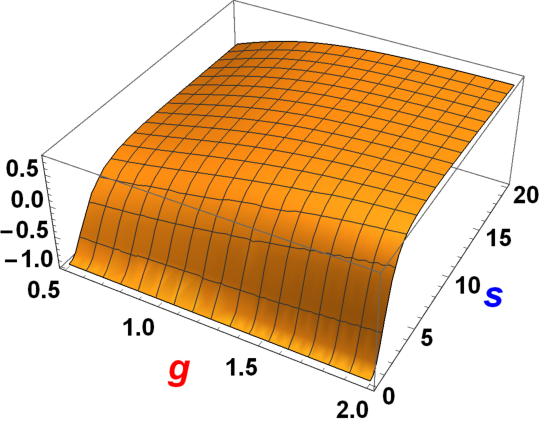}
\includegraphics[width=0.3\textwidth]{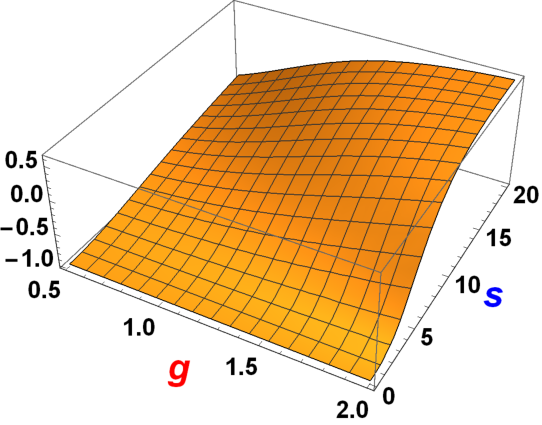}
\includegraphics[width=0.3\textwidth]{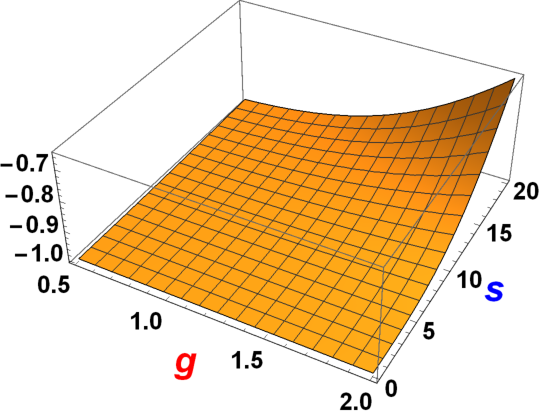}

\vspace{0.5cm}
\includegraphics[width=0.3\textwidth]{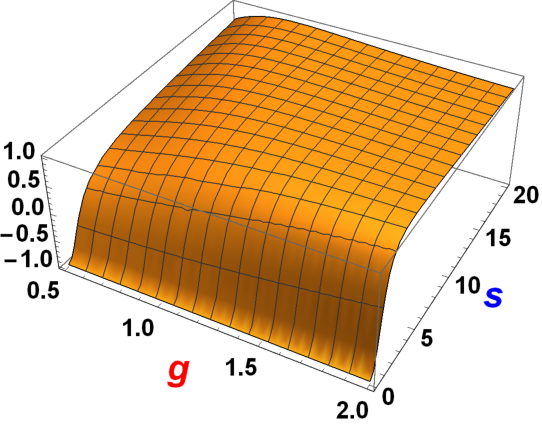}
\includegraphics[width=0.3\textwidth]{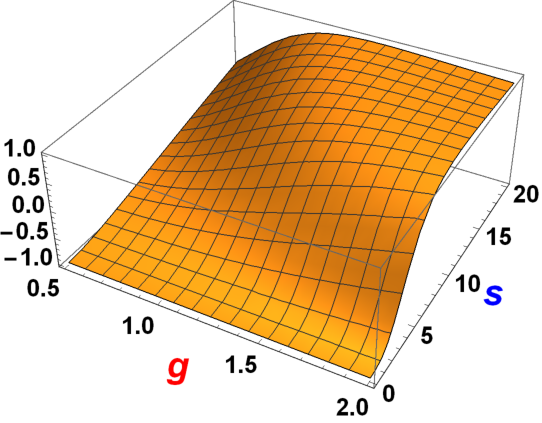}
\includegraphics[width=0.3\textwidth]{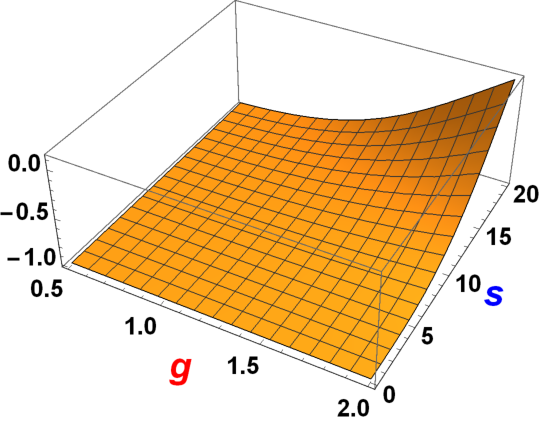}
 \parbox[t]{0.9\textwidth}{\caption{Asymmetry $A_{ll}$ as defined by Eq.\,(\ref{eq:Aij}) at the same conditions as in Fig.\,2 for unpolarized
 cross section.}\label{fig.3}}
\end{figure}

\subsection{Transverse polarization of beams in the reaction plane}

As one can see from (\ref{eq:S1S2}), polarization 4-vector $S_1^t$  in the rest frame of the electron has only space component which (for $\lambda_1=1)$ is $S_1 =(0, {\bf n}_{1t})$ and in this system
${\bf n}_{1t}\cdot{\bf p}_2 =0.$ Transition to c.m.s. does not change $S_1^t$, and because the same is valid for $S_2$
(with the change ${\bf n}_{1t}\rightleftarrows{\bf n}_{2t}, \ {\bf p}_2\rightleftarrows {\bf p}_1$), it is possible ${\bf n}_{1t}=\pm {\bf n}_{2t}$ only. The same is valid for the normal 4-vectors $S_1^n$ and $S_2^n$.

The $|M_h^{tt}|^2$, when both, the electron and positron are transversely polarized
in the reaction plane along the same direction can be written as

\begin{equation}\label{eq:Mhtt}
\left|M_{3/2}^{+tt} \right|^2 =  e^4 \,\big[ 2\, g_1^4 \,T_{11}(s,t) + \frac{1}{3}\, g_1^2 \,g_2^2 \,T_{12}(s,t) + \frac{1}{36}\, g_2^4 \,T_{22}(s,t)\big]
\end{equation}
$$T_{11}(s,t) = \frac{2 t^2 (s+t)^2}{(1-t)(1+s+t)},$$
$$T_{12}(s,t) = \frac{(s+t)[s^3-4st^3+t^4-5s^2t(1+t)]}{(1+s+t)^2} -\frac{t\,[8st^3+t^4+6s^3(1+t)+s^2t(5+13t)]}{(1-t)^2}$$
$$+\frac{s\,[s^3+9t^4+2st^2(1+9t)+s^2t(2+9t)]}{(1-t)(1+s+t)},$$
$$T_{22}(s,t)=\frac{3\,(s+t)^2\,[s^2t(t-2) +st^2(t-3)+t^4 +s^3(1+3t)]}{(1+s+t)^2}$$
$$+\frac{3\,t^2\,[t^4+6s^3(1+t) + st^2(3+5t)+2s^2t(5t+4)]}{(1-t)^2}$$
$$-\frac{2\,t(s+t)[s^3(6-12t)+5t^4 -s^2t(9+7t)+st^2(10t-9)]}{(1-t)(1+s+t)}.$$

The QED-HL interference in this case reads
\begin{equation}\label{eq:MhMett}
2\,Re\big[M_h\,M_e^{*}\big] = \frac{2\,t(s+t}{3(1-t)(1+s+t)}[-12\,g_1^2\,(2+s)+ g_2^2[3s^2t+8t^2+s(2+8t)].
\end{equation}

The corresponding expressions for angular distribution read
$$T_{11}(s,\theta)=\frac{s^4\,\sin^4{\theta}}{2[(2+s)^2-s^2\,\cos^2{\theta}]},$$
$$T_{12}(s,\theta) =\frac{s^4(1+\cos{\theta})[56-11s+4(3s-10)\cos{\theta}+s(10\cos^2{\theta}-12\cos^3{\theta}+\cos^4{\theta})]}{8\,(2+s+s\cos{\theta})^2}$$
$$+\frac{s^4[8+9s+(8-18s)\,\cos^2{\theta} +9s\,\cos^4{\theta}]}{4[(2+s)^2-s^2\cos^2{\theta}]}$$
$$+\frac{s^4(1-\cos{\theta})[56-11s+(40-12s)\cos{\theta} +s(10\,\cos^2{\theta}+12\,\cos^3{\theta}+\cos^4{\theta})]}{8(2+s-s\cos{\theta})^2},$$
$$T_{22}(s,\theta)=\frac{s^5\,\sin^2{\theta}\,[132+53s -36\cos^2{\theta}-58\cos^2{\theta}+5s\cos^4{\theta}]}{8[(2+s)^2-s^2\cos^2{\theta}}$$
$$+\frac{3s^5(1+\cos{\theta})^2\,[44-17s +(6s-40)\cos{\theta}+4(3+4s)\cos^2{\theta}-6s\cos^3{\theta}+s\cos^4{\theta}]}{16(2+s+s\cos{\theta})^2}$$
$$+\frac{3s^5(1-\cos{\theta})^2[44-17s+40\cos{\theta} +12\cos^2{\theta}+s(-6\cos{\theta}+16\cos^2{\theta}+6\cos^3{\theta}+\cos^{\theta})]}{16(2+s-s\cos{\theta})^2}.$$

The interference with the QED amplitude is
\begin{equation}\label{eq:MhMetttheta}
2\,Re\,\big[M_h\,M_e^{*}\big]^{tt} = \frac{2 s^2\sin^2{\theta}}{3[(2+s)^2-s^2\cos^2{\theta}]}\big[12\,g_1^2(2+s) - g_2^2s(2+s+2s\cos^2{\theta})\big].
\end{equation}

Asymmetry caused by the transverse polarizations of beams along the same direction is shown in Fig.\,4. It changes sign if directions of polarizations are opposite.

\begin{figure}
\centering
\includegraphics[width=0.3\textwidth]{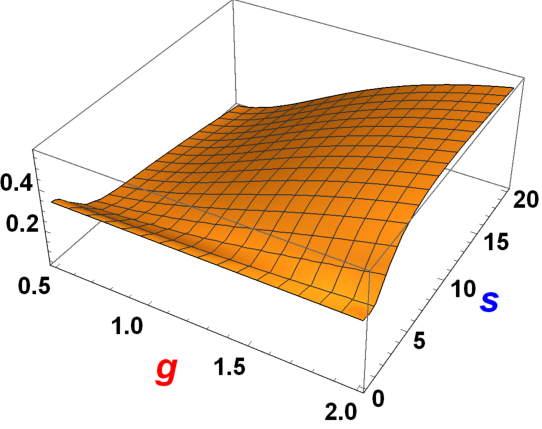}
\includegraphics[width=0.3\textwidth]{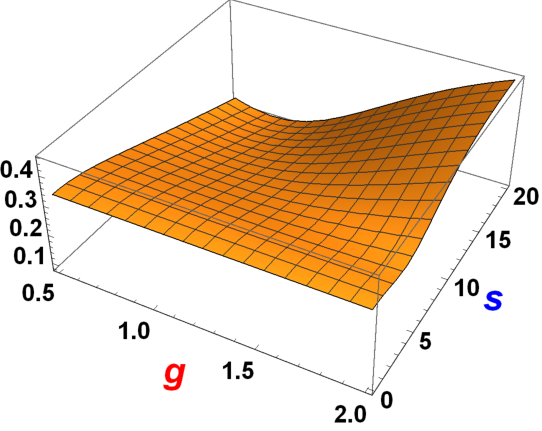}
\includegraphics[width=0.3\textwidth]{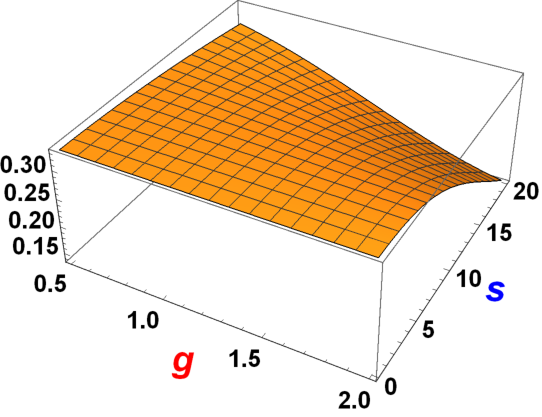}

\vspace{0.5cm}
\includegraphics[width=0.3\textwidth]{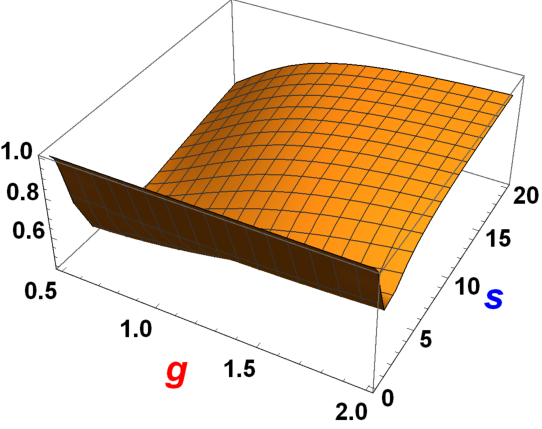}
\includegraphics[width=0.3\textwidth]{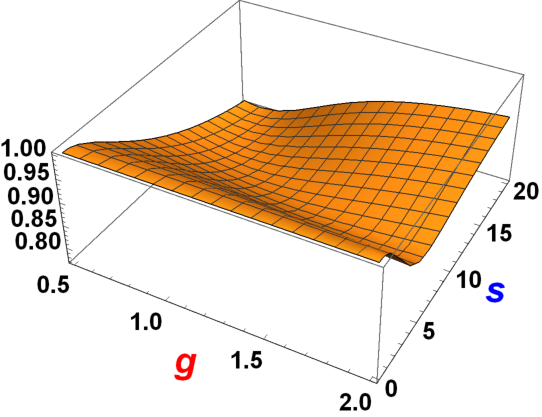}
\includegraphics[width=0.3\textwidth]{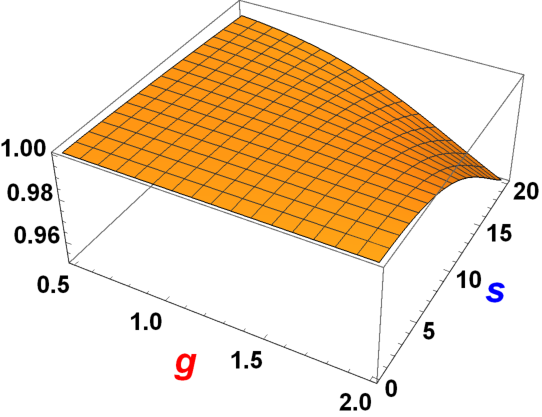}
 \parbox[t]{0.9\textwidth}{\caption{The same as in Fig.\,3 but for asymmetry $A_{tt}.$}\label{fig.4}}
\end{figure}

\subsection{longitudinal-transverse correlation}

If the positron beam is polarized longitudinally and the electron beam $-$ transversally (or vise versa), the corresponding spin-dependent
contribution is absent in the Born approximation. But contribution caused by the intermediate 3/2-spin heavy lepton is nonzero in this case and has a form
\begin{equation}\label{eq:Mhplet}
\left|M_h^{+lt} \right|^2 = (-st)\sqrt{\frac{s+t}{-st}}\Big(2g_1^3\,g_2\,P^{lt}_{31}(s,t) + g_1^2\,g_2^2\,P^{lt}_{22}(s,t)
+\frac{1}{3}g_1\,g_2^3\,P^{lt}_{13}(s,t) +\frac{1}{3}g_2^4\,P^{lt}_{04}(s,t)\Big),
\end{equation}
$$P^{lt}_{31}(s,t)=(1+2s+2s^2)\Big[\frac{1}{(1+s+t)^2} - \frac{1}{(1-t)^2}\Big] -(4+5s+s^2)\Big[\frac{1}{1+s+t} - \frac{1}{1-t}\Big] -2(s+2t),$$
$$P^{lt}_{22}(s,t) = -(1+2s+2s^2)\Big[\frac{1}{(1+s+t)^2} - \frac{1}{(1-t)^2}\Big]+(2+3s+s^2)\Big[\frac{1}{1+s+t} - \frac{1}{1-t}\Big],$$
$$P^{lt}_{13}(s,t) = (2-s)(s+2t) -\Big[\frac{1}{(1+s+t)^2} - \frac{1}{(1-t)^2}\Big] +4\Big[\frac{1}{1+s+t} - \frac{1}{1-t}\Big], $$
$$P^{lt}_{04}(s,t) = -(2+s)(s+2t) +(1+3s+3s^2)\Big[\frac{1}{(1+s+t)^2} - \frac{1}{(1-t)^2}\Big] +(4+8s+5s^2)\Big[\frac{1}{1+s+t} - \frac{1}{1-t}\Big].$$
As one can see, term with $g_1^4$ is absent and terms appear with the odd powers of $g_1$ and $g_2.$
If contrary, the electron beam is polarized longitudinally and the positron beam transversally, we have
$$P^{tl}_{31}(s,t) = P^{lt}_{31}(s,t), \ P^{tl}_{13}(s,t) = P^{lt}_{13}(s,t), \ P^{tl}_{22}(s,t) = -P^{lt}_{22}(s,t), \ P^{tl}_{04}(s,t) = -P^{lt}_{04}(s,t)$$
(terms with even powers $g_1$ and $_2$ change sign).

The QED-HL interference is also nonzero and reads
\begin{equation}\label{eq:MhMelt}
2Re\big[M_h\,M_e^{*}\big]^{lt}=2e^4s^2\sqrt{\frac{-t(s+t)}{s}}\Big[2g_1g_2\Big(\frac{1}{1+s+t} - \frac{1}{1-t}\Big) +g_2^2\Big(\frac{s+t}{1+s+t} + \frac{s+t}{1-t}\Big)\Big].
\end{equation}
In the case when the positron beam is polarized transversally and the electron one longitudinally, term with $g_1\,g_2$ change sign.

Taking into account that
$$(-st)\sqrt{\frac{s+t}{-st}} =\frac{s^{3/2}}{2}\sin{\theta},$$
one can rewrite corresponding the $(s,\,\theta)-$distribution
$$P^{lt}_{31}(s,\,\theta)=2s\cos{\theta}\Big[-1+ \frac{2(4+5s+s^2)}{(2+s)^2-s^2\cos^2{\theta}}-\frac{8(2+s)(1+2s+2s^2)}{[(2+s)^2-s^2\cos^2{\theta}]^2}\Big],$$
$$P^{lt}_{22}(s,\,\theta) =4s\cos{\theta}\Big[-\frac{2+3s+s^2}{(2+s)^2-s^2\cos^2{\theta}}+\frac{4(2+s)(1+2s+2s^2)}{[(2+s)^2-s^2\cos^2{\theta}]^2}\Big],$$
$$P^{lt}_{13}(s,\,\theta) = s\cos{\theta}\Big[2-s -\frac{16}{(2+s)^2-s^2\cos^2{\theta}} + \frac{16(2+s)}{[(2+s)^2-s^2\cos^2{\theta}]^2}\Big],$$
$$P^{lt}_{04}(s,\,\theta) = s\cos{\theta}\Big[-2-s -\frac{4(4+8s+5s^2)}{(2+s)^2-s^2\cos^2{\theta}} - \frac{16(2+s)(1+3s+3s^2)}{[(2+s)^2-s^2\cos^2{\theta}]^2}\Big].$$

The QED-HL  interference for this distribution is
\begin{equation}\label{eq:MhMelttheta}
2\,Re\,\big[M_{3/2}^+\,M^{e*}\big]^{lt} = 4\,s^{5/2}\sin{\theta}\,\cos{\theta}\,\frac{-2g_1g_2\,s +g_2^2(2+s)}{(2+s)^2-s^2\cos^2{\theta}}.
\end{equation}
It is clear that corresponding asymmetries go to zero if $\theta$ go to 90$^o.$ They are shown in Figs.\,5,\,6.
\begin{figure}
\centering
\includegraphics[width=0.3\textwidth]{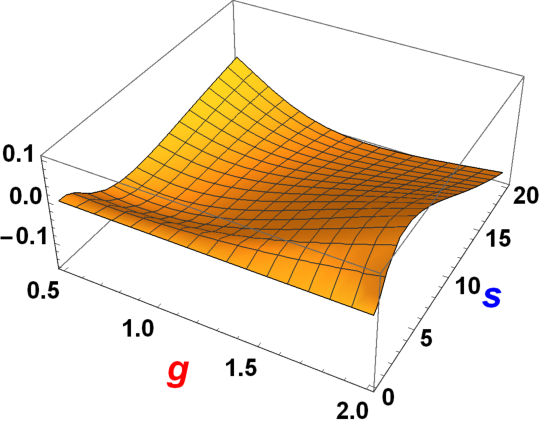}
\includegraphics[width=0.3\textwidth]{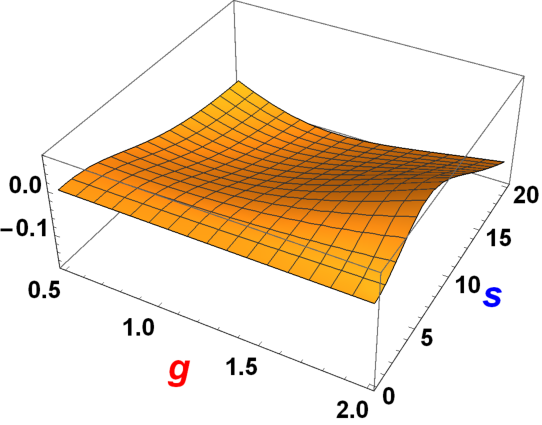}
\includegraphics[width=0.3\textwidth]{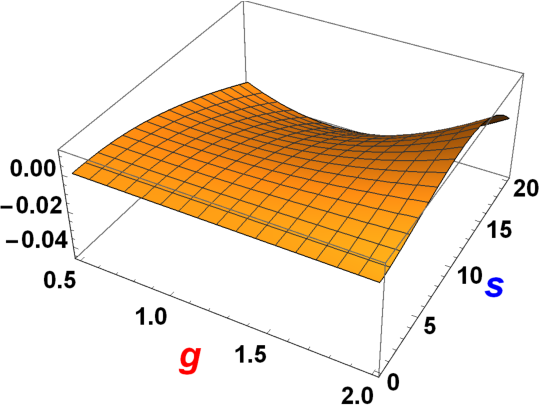}

\vspace{0.5cm}
\includegraphics[width=0.3\textwidth]{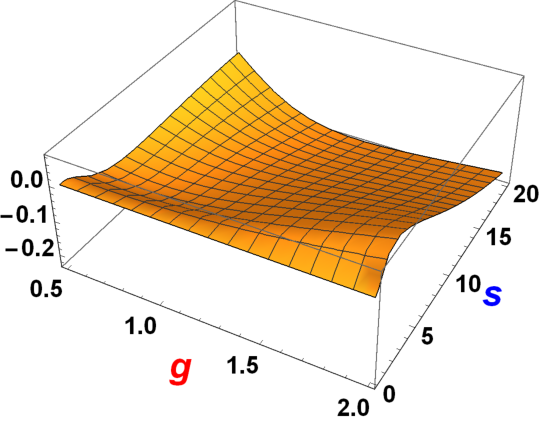}
\includegraphics[width=0.3\textwidth]{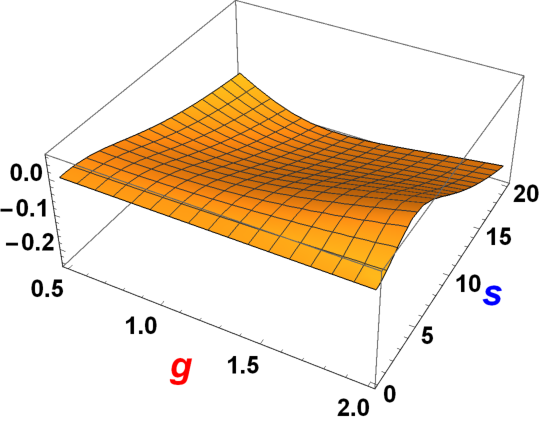}
\includegraphics[width=0.3\textwidth]{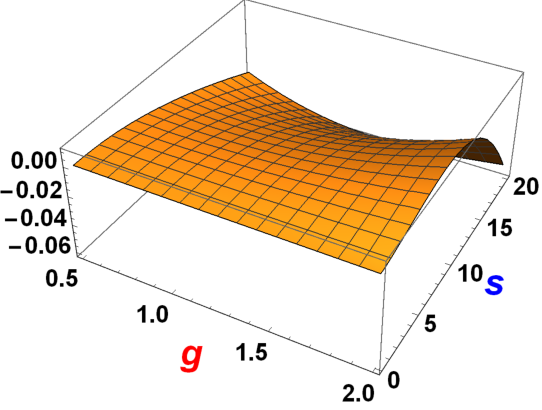}
 \parbox[t]{0.9\textwidth}{\caption{Asymmetry $A_{lt}$ at photon angle $\theta =45^o$ (the upper row); $\theta=60^0$ (lower row)
 and the values of constants  $g_1$ as in Fig.\,2. It arises due to intermediate 3/2-spin heavy lepton and goes to zero at $\theta \to \pi/2.$}\label{fig.5}}
\end{figure}

\begin{figure}
\centering
\includegraphics[width=0.3\textwidth]{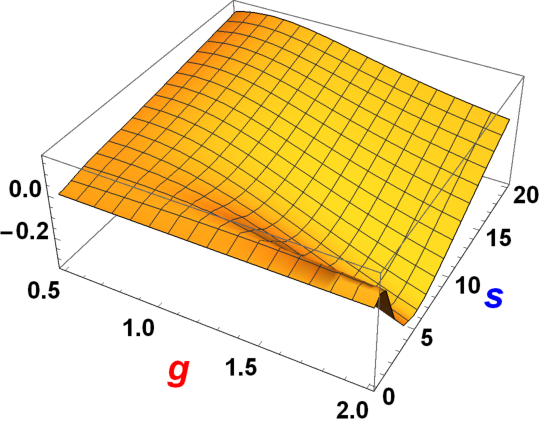}
\includegraphics[width=0.3\textwidth]{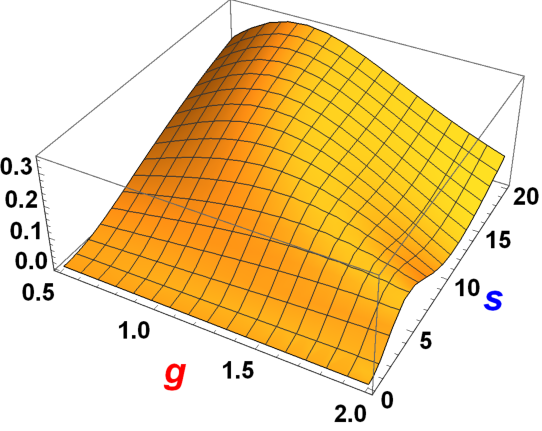}
\includegraphics[width=0.3\textwidth]{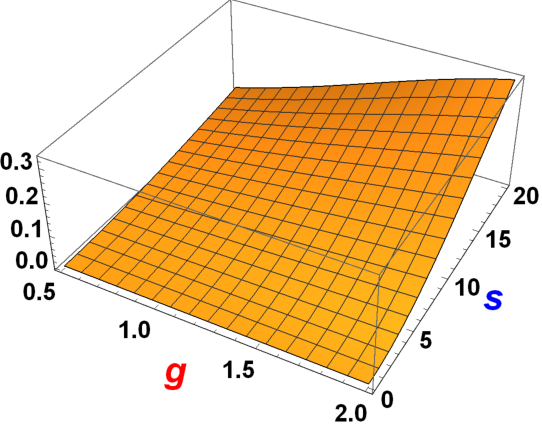}

\vspace{0.5cm}
\includegraphics[width=0.3\textwidth]{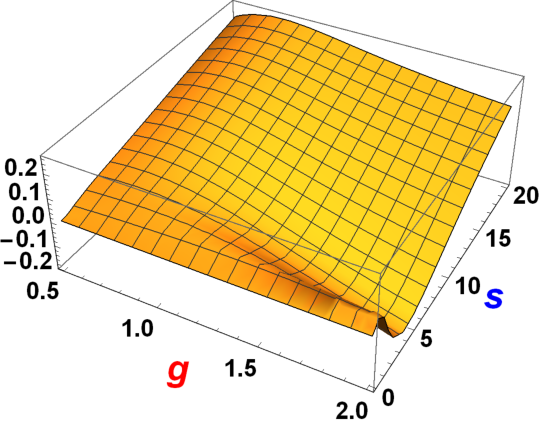}
\includegraphics[width=0.3\textwidth]{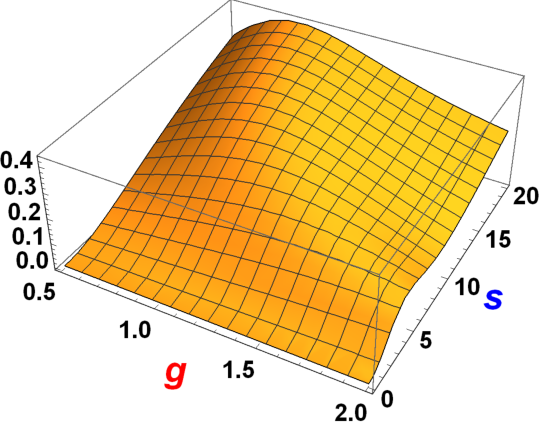}
\includegraphics[width=0.3\textwidth]{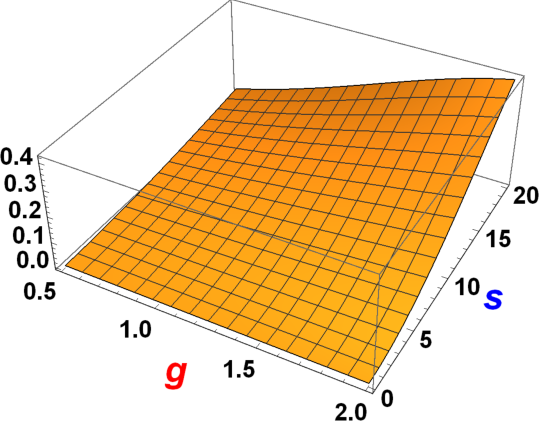}
 \parbox[t]{0.9\textwidth}{\caption{The same as in Fig.\,5 but for asymmetry $A_{tl}.$ }\label{fig.6}}
\end{figure}

\subsection{normal polarizations }

As it noted above, at $\eta_1=\eta_2 =1,$ the 4-vectors of normal polarization are $S_1^n = (0,\,{\bf n_{1n}})$ and $S_2^n = (0,\,{\bf n_{2n}})$ with ${\bf n_{1n}}=\pm {\bf n_{2n}}.$
If both beams are polarized perpendicularly to the reaction plane in the same direction we have
\begin{equation}\label{eq:Mhnn}
|M_h|^2 = 2g_1^4\,N_{11}(s,t) +\frac{1}{3}g_1^2\,g_2^2\,N_{12}(s,t) + \frac{1}{36}N_{22}(s,t),
\end{equation}
$$N_{11}(s,t)= \frac{2t^2(s+t)^2}{(1-t)(1+s+t)}, $$
$$N_{12}(s,t)= \frac{[s^3-4st^3+t^4-5s^2t(1+t)](s+t)}{(1+s+t)^2} - \frac{t[8st^3+t^4+6s^3(1+t)+s^2t(5+13t)]}{(1-t)^2}$$
$$+\frac{s\,[s^3+9t^4+2st^2(1+9t)+s^2t(2+9t)]}{(1-t)(1+s+t)},$$
$$N_{22}(s,t)=\frac{3(s+t)^2[s^2t(t-2)+st^2(t-3)+t^4+s^3(1+3t)]}{(1+s+t)^2}$$
$$+\frac{3t^2[t^4+6s^3(1+t)+st^2(3+5t)+2s^2t(4+5t)]}{(1-t)^2}$$
$$-\frac{2t[s^3(6-12t)+5t^4-s^2t(9+7t)+st^2(10t-9)]}{(1-t)(1+s+t)}.$$

The QED-HL interference in this case coincides with (\ref{eq:MhMett}).

$$N_{11}(s,\theta)= -2(1+s) +\frac{s^2\,\sin^2{\theta}}{2}+ \frac{8(1+s)}{(2+s)^2-s^2\cos^2{\theta}},$$
$$N_{12}(s,\theta) = -8-6s-10s^2-\frac{s^2}{2} +\frac{s^2(s-2)\cos^2{\theta}}{2} +\frac{4(10+18s+22s^2+12s^3-s^4)}{(2+s)^2-s^2\cos^2{\theta}}$$
$$-\frac{8\,(1+2s+2s^2)[(2+s)^2+s^2\cos^2{\theta}]}{[(2+s)^2-s^2\cos^2{\theta}]^2},$$
$$N_{22}(s,\theta) = 40+92s+63s^2-34s^3+\frac{s^4}{4}+\frac{s^2(28+52s-5s^2)\cos^2{\theta}}{4} +s^4\cos^4{\theta}$$
$$-\frac{4(46+144s+177s^2+52s^3-18s^4)}{(2+s)^2-s^2\cos^2{\theta}} + \frac{24(1+2s+2s^2)((2+s)^2+s^2\cos^2{\theta})}{[(2+s)^2-s^2\cos^2{\theta}]^2}.$$

As well as longitudinal and transverse asymmetries the normal one does not contain the odd powers of $g_1$ and $g_2.$ It plotted in Fig.\,7.

\begin{figure}
\centering
\includegraphics[width=0.3\textwidth]{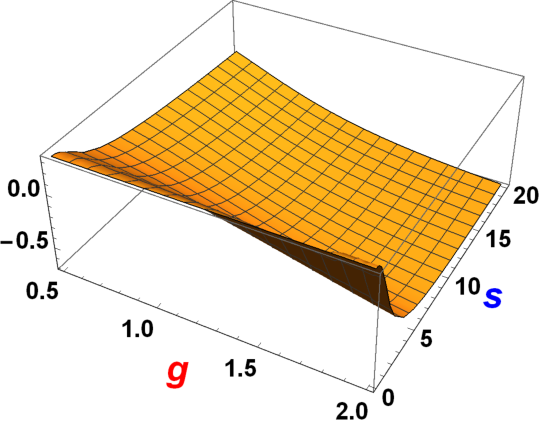}
\includegraphics[width=0.3\textwidth]{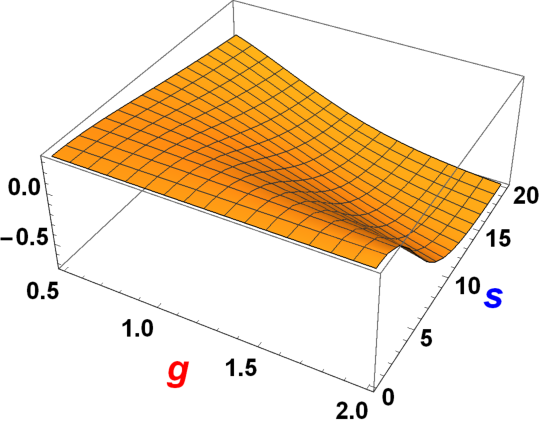}
\includegraphics[width=0.3\textwidth]{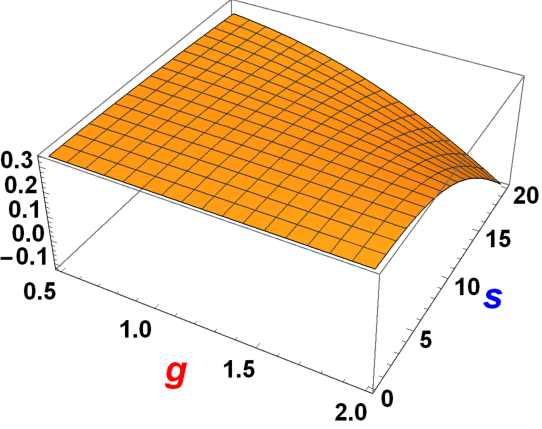}

\vspace{0.5cm}
\includegraphics[width=0.3\textwidth]{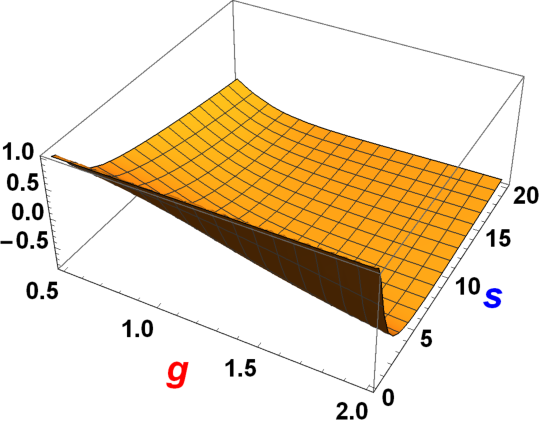}
\includegraphics[width=0.3\textwidth]{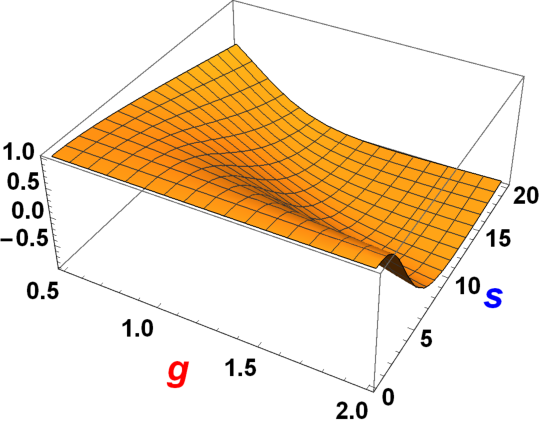}
\includegraphics[width=0.3\textwidth]{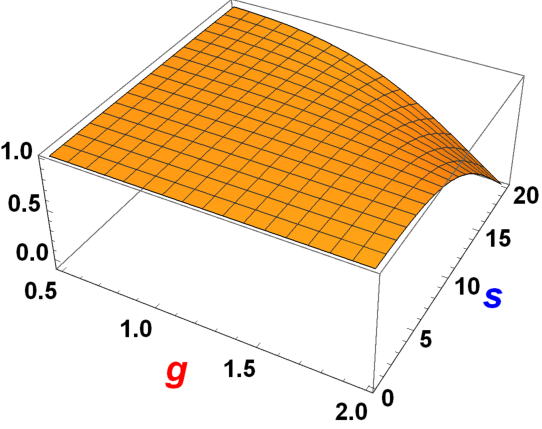}
 \parbox[t]{0.9\textwidth}{\caption{the same asin Fig.\,3 but for asymmetry $A_{nn}$.}\label{fig.7}}
\end{figure}

\section{Discussion and conclusion}

Manifestation of the 3/2-spin heavy lepton $h^\pm$ has been considered in large-angle $e^+\,e^- \to \gamma\,\gamma$ reaction at high energies. We suggest that lepton $h^\pm$ can
interact with QED sector of the SM and appears as a virtual intermediate state in Feynmann diagrams, changing the intermediate electrons in $t$ and $u$ channels of this reaction.
Phenomenological Lagrangian of the $h^\pm\,e^\pm\,\gamma$ interaction chosen similarly to one describing decay of the $\Delta$ isobar into electron and photon. It contains two unknown real constants which are dimensionless by dividing by the parameter with the dimension of mass, and we take this parameter as the heavy lepton mass $M$.

At energies $\sqrt{s}\approx$\,100\,GeV the born differential section (\ref{eq:Born}) is about 2\,pb and decreases with increasing energy as $s^{-1}$. Electroweak radiative corrections (EWRC) to the total cross section at such energies consist of about few percent and the positive QED share partially compensated by a negative weak correction \cite{CarloniCalame:2019dom, Bondarenko:2022xmc}.
Contribution due to 3/2-spin heavy lepton increases very quickly with the energy and therefore can, in principle, be competitive with EWRC at ILC and CLIC colliders.

Our finding is that at $g_1=1$ and $M=$\,100\,GeV contribution due to HL and QED-HL interference dominates already at $\sqrt{s}=$\,50\,GeV and such situation seems implausible. At $g_1=0.35$ it is of the order born cross section what is unrealistic too, and if $g_1=0.1$ it can be on the level of EWRC. Very interesting is the possibility to probe effects due to heavy lepton by measuring the longitudinal-transverse asymmetries $A_{lt}$ and $A_{tl}$ Figs.\,5,\,6 where the pure QED mechanism does not contribute.

{\bf Acknowledgment}

The authors acknowledge support by the National Academy of Sciences of Ukraine via the program
“Participation in the international projects in high energy and nuclear physics” (project no. 0121U111693).

\end{document}